\documentclass[10pt,conference]{IEEEtran}
\usepackage{cite}
\usepackage{amsmath,amssymb,amsfonts}
\usepackage{algorithmic}
\usepackage{graphicx}
\usepackage{textcomp}
\usepackage{xcolor}
\usepackage[hyphens]{url}
\usepackage{fancyhdr}
\usepackage{hyperref}

\newcommand{\hpcayear}{2027}

\def\hpcacameraready{} 

\usepackage{url}
\usepackage{amsmath,amsfonts}
\usepackage{graphicx}
\usepackage{textcomp}
\usepackage{subfig}
\usepackage{color}
\usepackage[dvipsnames,svgnames,table]{xcolor}
\usepackage{multirow}
\usepackage{lipsum}
\usepackage{caption}
\usepackage{wrapfig}
\usepackage{makecell}
\usepackage{xspace}
\usepackage{pifont}

\usepackage[linecolor=DarkGreen,backgroundcolor=DarkGreen!25,bordercolor=DarkGreen]{todonotes}
\usepackage{lipsum}
\usepackage{caption}
\usepackage{booktabs}
\usepackage{lipsum}
\usepackage{adjustbox}
\usepackage{bm}
\usepackage{soul}
\usepackage{listings}
\usepackage{xcolor}
\usepackage[noend, ruled, vlined, linesnumbered, noline, boxed]{algorithm2e}
\usepackage{threeparttable}
\usepackage{tikz}

\DeclareRobustCommand{\circled}[1]{%
  \tikz[baseline=(char.base)]{
    \node[shape=circle,fill=black,inner sep=1pt] (char)
    {\color{white}\normalfont\sffamily\scriptsize #1};}}

\newcommand{\name}{\textit{A-Graph}\xspace}
\newcommand{\thiswork}{\textit{Archx}\xspace}
\usepackage[table]{xcolor}
\definecolor{tabgreen}{HTML}{DDF2DA}
\newcommand{\tabemph}[1]{%
    \cellcolor{tabgreen}\textit{#1}%
}

\makeatletter
\newcommand{\minisection}[1]{%
  \par\smallskip 
  \noindent\textbf{#1}~
  \ignorespaces
}
\makeatother

\title{\thetitle}

\newcommand\hpcaauthors{%
Daniel Price$^{1}$,
Prabhu Vellaisamy$^{2}$,
Patricia Gonzalez$^{3}$,
George Michelogiannakis$^{3}$,
John P. Shen$^{2}$,
Di Wu$^{1}$}

\newcommand\hpcaaffiliation{%
$^{1}$Department of ECE, University of Central Florida, Orlando, FL, USA\\
$^{2}$Department of ECE, Carnegie Mellon University, Pittsburgh, PA, USA\\
$^{3}$Lawrence Berkeley National Laboratory, Berkeley, CA, USA}

\newcommand\hpcaemail{%
daniel.price@ucf.edu,
pvellais@andrew.cmu.edu,
lg4er@lbl.gov,
mihelog@lbl.gov,
jpshen@cmu.edu,
di.wu@ucf.edu}

\newcommand{\thetitle}{\name: A Unified Graph Representation for Cross-Stack Cost Modeling}

\author{
  \ifdefined\hpcacameraready
    \IEEEauthorblockN{\hpcaauthors{}}
      \IEEEauthorblockA{
        \hpcaaffiliation{} \\
        \hpcaemail{}
      }
  \else
    \IEEEauthorblockN{\normalsize{HPCA \hpcayear{} Submission
      \textbf{\#\hpcasubmissionnumber{}}} \\
      \IEEEauthorblockA{
        Confidential Draft \\
        Do NOT Distribute!!
      }
    }
  \fi 
}

\fancypagestyle{camerareadyfirstpage}{%
  \fancyhead{}
  \renewcommand{\headrulewidth}{0pt}
  \fancyhead[C]{
    \ifdefined\aeopen
    \parbox[][12mm][t]{13.5cm}{\hpcayear{} IEEE International Symposium on High-Performance Computer Architecture (HPCA)}    
    \else
      \ifdefined\aereviewed
      \parbox[][12mm][t]{13.5cm}{\hpcayear{} IEEE International Symposium on High-Performance Computer Architecture (HPCA)}
      \else
      \ifdefined\aereproduced
      \parbox[][12mm][t]{13.5cm}{\hpcayear{} IEEE International Symposium on High-Performance Computer Architecture (HPCA)}
      \else
      \parbox[][0mm][t]{13.5cm}{\hpcayear{} IEEE International Symposium on High-Performance Computer Architecture (HPCA)}
    \fi 
    \fi 
    \fi 
    \ifdefined\aeopen 
      \includegraphics[width=12mm,height=12mm]{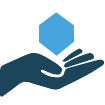}
    \fi 
    \ifdefined\aereviewed
      \includegraphics[width=12mm,height=12mm]{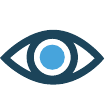}
    \fi 
    \ifdefined\aereproduced
      \includegraphics[width=12mm,height=12mm]{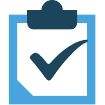}
    \fi
  }
  \fancyfoot[C]{}
}
\renewcommand{\headrulewidth}{0pt}

\begin{document}
\maketitle

\ifdefined\hpcacameraready 
  \thispagestyle{camerareadyfirstpage}
  \pagestyle{empty}
\else
  \thispagestyle{plain}
  \pagestyle{plain}
\fi

\newcommand{\hpcaheight}{0mm}
\ifdefined\eaopen
\renewcommand{\hpcaheight}{12mm}
\fi

\makeatletter
\fancypagestyle{camerareadyfirstpage}{%
  \fancyhf{}
  \renewcommand{\headrulewidth}{0pt}
  \fancyfoot[C]{}
}
\makeatother


\begin{abstract}

Emerging systems, such as superconducting and neuromorphic, are driving a new era of computing with increasingly diverging design requirements compared to conventional systems.
These rapidly evolving design spaces require agile development and early design-space exploration (DSE) to identify promising performance-cost tradeoffs.
While conventional systems benefit from mature toolchains, emerging systems lack comparable support.
Existing cost estimators target specific stacks, while hardware-generation frameworks presume mature compilers and EDA flows absent in such emerging technologies.

To fill this gap, we introduce Architecture-Graph (\name), a graph that unifies the cross-stack system representation, realized through \thiswork for e\underline{x}ploring the large \underline{arch}itecture design space.
\name and \thiswork form a cross-stack \emph{cost-modeling} framework that separates cost aggregation from performance models, enabling tightly coupled cost co-simulation.

\name and \thiswork facilitate \textit{usability} in three key aspects:
\circled{1}~\name supports cross-stack system representation and metric definition, enabling \textit{flexibility}.
\circled{2} \thiswork automatically generates and sweeps design points under user constraints, improving \textit{programmability}.
\circled{3} \thiswork adopts scope-based metric retrieval and aggregation exposure, reporting cost at any user-defined hierarchy, enhancing \textit{explainability}.

Case studies spanning emerging systems (superconducting, neuromorphic) and conventional architectures (RISC-V, GPU) demonstrate high fidelity and showcase usability.
Overall, \name and \thiswork unlock cross-stack cost modeling for rapid and accurate DSE.

\end{abstract}
\section{Introduction}
\label{sec:Introduction}




Computer architecture is entering a golden age~\cite{comparch_golden}, with new computing systems flourishing.
Beyond traditional CMOS, more exotic circuit technologies are emerging, such as memristors~\cite{thomas2013memristor, chanthbouala2012ferroelectric, li2018analogue, chua2003memristor, yao2020fully, williams2008we}, superconducting~\cite{allen1983theory, clarke2008superconducting, savitskii2012superconducting, superconducting_cnn, michelogiannakis2025unconventional, superconducting_fir, Temporal_Logic_Superconducting}, photonics~\cite{dong2024partial, shastri2021photonics, drake1986photonic, antonik2019human, rios2019memory, van2017advances}, quantum~\cite{o2007optical, schrader2004neutral, isenhower2010demonstration, rudolph2017optimistic, blatt2012quantum, pino2021demonstration}, and more (even mushroom~\cite{larocco2025sustainable}).
Each technology instigates new architectural organizations: memristors ignite neuromorphic computing~\cite{liu2015spiking, 10.1145/3316781.3317870, adam20163, ankit2017resparc}, while superconducting and photonic devices motivate new architectures for artificial intelligence (AI)~\cite{superconducting_cnn, lin2018all, ahmed2025universal} and quantum computing~\cite{ding2025high, psiquantum2025manufacturable, clarke2008superconducting}.
New applications naturally arise, such as AI for protein structure prediction~\cite{jumper2021highly}, physics simulation~\cite{hennigh2021nvidia}, cryogenic sensing~\cite{cardani2017new, stolz2021superconducting}, and more.


\begin{table*}[!t]
    \centering
    \begin{adjustbox}{max width=\linewidth}
        \begin{threeparttable}
            \caption{Comparing simulators for hardware cost estimation. Green cells indicate the best usability in each category.}
            \label{tab:prior_works}
            \setlength{\tabcolsep}{3pt}
            \begin{tabular}{cccccccccc}
                \toprule
                \multirow{2}{*}[-0.6ex]{\textbf{\small Design}} & \multicolumn{2}{c}{\multirow{2}{*}[-0.6ex]{\textbf{\small Goal}}} & \multicolumn{3}{c}{\textbf{\small Flexibility}} & \multicolumn{2}{c}{\textbf{\small Programmability}} & \multicolumn{2}{c}{\textbf{\small Explainability}} \\
                \cmidrule(lr){4-6}\cmidrule(lr){7-8}\cmidrule(lr){9-10}
                                           & \multicolumn{2}{c}{}   & \textbf{Stack}        & \textbf{Metric}        & \textbf{Target}                & \textbf{Expertise}      & \textbf{Sweeping}   & \textbf{Scope}   & \textbf{Aggregation} \\
                \midrule
                Charm~\cite{charm}         & Pre-RTL  & Perf / cost & Arch                  & \tabemph{User-defined} & Analytical                     & \tabemph{Arch behavior} & \tabemph{Automated} & End-to-end       & Obscured \\
                Aladdin~\cite{alladin}     & Pre-RTL  & Perf / cost & \tabemph{Cross-stack} & PPA                    & CMOS                           & C                       & Partially auto      & Component        & Obscured \\
                SNSv2~\cite{snsv2}         & Post-RTL & Perf / cost & Arch/Circuit          & PPA                    & CMOS                           & RTL                     & Manual              & End-to-end       & Obscured \\
                DSAGEN~\cite{dsagen}       & Post-RTL & Chip gen    & \tabemph{Cross-stack} & PPA                    & CMOS Acc              & C \& \#pragma           & Partially auto      & End-to-end       & Obscured \\
                OverGen~\cite{overgen}     & Post-RTL & Chip gen    & \tabemph{Cross-stack} & PPA                    & CMOS FPGAs                     & C \& \#pragma           & Partially auto      & End-to-end       & Obscured \\
                TNNGen~\cite{tnngen}       & Post-RTL & Chip gen    & \tabemph{Cross-stack} & PPA                    & CMOS TNN                       & C \& \#pragma           & Manual              & End-to-end       & Obscured \\
                Accelergy~\cite{accelergy} & Pre-RTL  & Cost        & \tabemph{Cross-stack} & Energy                 & Timeloop~\cite{timeloop_paper} & \tabemph{Arch behavior} & Manual              & Component        & Obscured \\
                \midrule
                \thiswork (ours)           & Pre-RTL  & Cost        & \tabemph{Cross-stack} & \tabemph{User-defined} & \tabemph{User-defined}         & \tabemph{Arch behavior} & \tabemph{Automated} & \tabemph{Scoped} & \tabemph{Exposed} \\
                \bottomrule
            \end{tabular}
            \begin{tablenotes}[para]
                \footnotesize
                \item PPA: Performance, power and area.
                \item TNN: Temporal neural network (Neuromorphic).
            \end{tablenotes}
        \end{threeparttable}
    \end{adjustbox}
\end{table*}
Within these new systems, a wide range of simulators exist.
Circuit simulators support emerging paradigms such as memristors~\cite{yakopcic2011memristor}, superconducting~\cite{delport2019josim}, photonics~\cite{chan2010phoenixsim}, etc.
Architecture simulators explore computer architecture and memory design~\cite{charm, binkert2011gem5, rodrigues2011structural, alladin, rosenfeld2011dramsim2, cacti7, xu2018pimsim}.
Application simulators focus on specific workloads, such as AI with general matrix multiplication (GEMM)~\cite{timeloop_paper, maestro_micro2019, cosa_paper, ASTRA_sim_v2}, visual computing~\cite{ma2023camj}, unary computing~\cite{2020isca_ugemm, 2022hpca_usystolic, carat_paper, mugi_paper}, neuromorphic computing~\cite{SuperNeuro_paper, yin2022sata}, robotics~\cite{nikiforov2023rose}, etc.
Cost simulators, e.g., Accelergy~\cite{accelergy}, target varying stacks for metric cost analysis.
Hardware-generation frameworks generate hardware directly from user specifications or programs~\cite{dsagen, overgen, tnngen}.

While these new systems promise substantial gains, evaluating their performance and cost within existing frameworks remains challenging.
Their design spaces are vast and rapidly evolving, lowering the efficacy of these specialized simulators in the near future and imposing \emph{usability challenges}.

\begin{figure}[!t]
  \centering
  \includegraphics{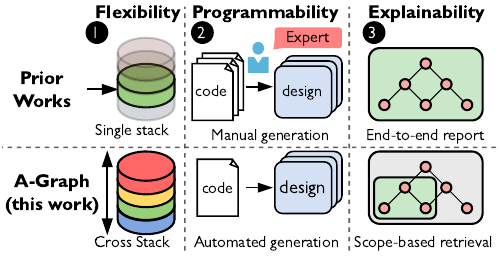}
  \caption{Usability challenges.}
  \label{fig:usability}
\end{figure}

\minisection{Usability challenges:}
\label{minisec:usability_challenges}
Figure~\ref{fig:usability} outlines three usability challenges we have identified where existing frameworks fall short.


\begin{itemize}
    \item \textbf{\textit{Flexibility}}:
    While few existing frameworks lack cross-stack support, most are specialized for a target stack.
    Moreover, they are often hardwired to specific metrics, limiting cross-stack co-design and hindering \textit{flexibility} as emerging systems arise.

    \item \textbf{\textit{Programmability}}: 
    DSE in existing frameworks often requires manually defining configurations.
    This can be time-consuming when exploring thousands of design points, and error-prone as singular mistakes can invalidate comparative results.
    Moreover, some existing works require high cross-stack expertise when designing in their frameworks.
    This lack of programmability hinders the rapid design process that emerging systems demand.
    
    \item \textbf{\textit{Explainability}}:
    Existing frameworks offer high fidelity, but often limit reported metrics to end-to-end results.
    Additionally, those with hierarchical reporting do not expose how these metrics are computed.
    Without explainability, understanding interactions in emerging systems with unknown design spaces is difficult.

\end{itemize}

Table~\ref{tab:prior_works} compares prior works along these axes of usability.
Each simulator is effective within its target stack(s), where specialized cost models and compilation frameworks lend themselves to high fidelity.
However, this specialization causes each to fall short in at least one aspect of usability.

By targeting specific stacks or metrics, these frameworks lack the flexibility to adapt to emerging systems.
Extending them would not only require modifying their performance models, but also redesigning their underlying cost models or compilation frameworks.
Even with modifications, existing frameworks would remain limited for DSE, as they lack the programmability and explainability required to efficiently explore and interpret vast design spaces.
Moreover, continuously redesigning frameworks as new systems emerge delays development, consuming effort that could otherwise be spent exploring and optimizing design decisions.
These deficiencies raise a key question: \textit{Can a single framework provide the flexibility, programmability, and explainability needed to enable cross-stack cost analysis and DSE for emerging systems?}

\minisection{Proposal:}
In response to this challenge, we introduce Architecture-Graph (\name), a graph that unifies the representations of system stacks.
Unlike prior works that are constrained to target stacks or predefined metrics, \name allows designers to customize their own performance models and cost metrics.
Represented in the same graph, performance models break down the full system into hierarchical events that ultimately link to their own cost.
This representation lets designers model target emerging systems without hard-wiring metric aggregation.
Moreover, designers can now select the simulation fidelity by providing the right performance models and cost metrics at the target granularity, where existing hardwired frameworks fail.

To unlock \name's full potential, we develop \thiswork, a cost-modeling framework that implements \name.
\thiswork consists of a front-end programming interface for configuration sweeping, which automatically generates and sweeps only valid design points in the vast design space.
Additionally, \thiswork provides scope-based metric retrieval and aggregation exposure, allowing designers to both evaluate system metrics at scopes of interest and reveal how such metrics are computed.
This programmability equips \name for the rapid DSE emerging systems require, while the explainability exposes system interactions within \name.


Together, \name and \thiswork form a cost-modeling framework that separates cost aggregation from performance models that drive it.
With this separation, representing a system only requires supplying its performance models and cost metrics, while cost aggregation stays fixed.
This yields tightly coupled cost co-simulation without rebuilding the framework, addressing the usability challenges above.

Here we outline how we position \name in the ecosystem.

\minisection{What \name/\thiswork is}
\begin{itemize}
    \item a unified representation for cross-stack cost evaluation.
    \item a decoupled framework of cost aggregation from user-supplied performance models and metrics.
    \item an automated approach for cost modeling towards DSE.
    \item a framework for scope-based metric retrieval and aggregation exposure.
\end{itemize}

\minisection{What \name/\thiswork is not}
\begin{itemize}
    \item a simulator with built-in performance models or metrics.
    \item a framework restricted to a specific stack or technology.
    \item an EDA or compiler toolchain.
\end{itemize}

The contributions of this work are as follows:

\begin{itemize}
    \item We introduce \name, a unified cross-stack graph representation that separates cost aggregation from user-supplied performance models and customizable metrics, enabling flexible cost evaluation for emerging systems.

    \item We develop \thiswork, a cost-modeling framework realizing \name, that automatically sweeps design points under user constraints, improving programmability, and exposes cost at user-defined scopes, enhancing explainability.

    \item We validate \name and \thiswork through case studies covering both emerging and conventional systems, across different technologies, architectures, and applications, achieving high fidelity.
\end{itemize}

This paper is organized as follows.
Section~\ref{sec:motivation} motivates \name.
Sections~\ref{sec:a-graph} and~\ref{sec:architecture} detail \name and \thiswork.
Sections~\ref{sec:experimental_setup} and~\ref{sec:case_study} present case studies.
Sections~\ref{sec:discussion} and~\ref{sec:conclusion} discuss and conclude this paper.

\section{Motivating Usability}
\label{sec:motivation}

Section~\ref{sec:Introduction} named three usability challenges that prevent existing frameworks from serving emerging systems: flexibility, programmability, and explainability, detailed in Table~\ref{tab:prior_works}.



\subsection{Flexibility}
Most existing frameworks are constrained to their target stack(s), limiting their flexibility to emerging systems.
Some of these frameworks are able to cross the stack within such constraints.
Accelergy can attribute cost only for AI tensor workloads by integrating with Timeloop~\cite{accelergy, timeloop_paper}.
Aladdin estimates cost directly from C~\cite{alladin}, while DSAGEN and OverGen generate RTL from C and compiler directives~\cite{dsagen, overgen}.
All three are confined to CMOS, while the latter two are further confined to spatial accelerators or FPGAs.
TNNGen is limited to neuromorphic workloads, generating RTL from PyTorch implementations~\cite{tnngen}.
Other works focus on one specific stack.
Charm only models architecture interactions~\cite{charm}, while SNSv2 estimates synthesis results from RTL~\cite{snsv2}.

Additionally, such frameworks hardwire metrics that are needed for their target stack(s).
Accelergy reports only energy, while most other works are limited to CMOS metrics, reporting only performance, power, and area (PPA).
A cost outside this built-in set cannot be expressed at all.
For example, area for superconducting logic is measured in Josephson junctions (JJ) rather than $\text{mm}^2$ for CMOS, which no existing framework can be extended to support.
Charm alone allows user-defined metrics, though only at an analytical level.

Mature EDA tools deliver precise PPA estimates, but each is similarly bound to a single technology and stack (e.g., CMOS at the circuit level) and to metrics that technology supports.
Emerging technologies such as superconducting and memristor logic lack such mature tools, though recent works begin to address this~\cite{edaq_superconducting, qiskit_metal, eda_memristor, atomic_memristor}.

Existing frameworks lack the flexibility that emerging systems require, motivating a framework that represents cross-stack system interactions with customizable metrics.

\subsection{Programmability}
\label{subsec:programmability}

Effective DSE requires rapidly generating and evaluating many configurations, which is time-consuming in existing frameworks.
Without constraining the design space to only valid configurations, time is wasted evaluating invalid designs. 
Assume a designer wants to explore general matrix multiplication (GEMM) on a systolic array.
This involves sweeping across the array size, bitwidth, dataflow, buffer size, GEMM dimensions (M, K, N), tiling, etc.
With only two options per knob, these eight knobs already give 256 combinations, a count that grows exponentially with every knob or configuration added.
Conversely, frameworks that do not include automation under constraints require the designer to manually write configurations, which is error-prone.
If a single mistake is made among thousands of combinations, then the comparative results are invalidated, and thus the conclusion.

Existing frameworks include varying levels of automation, but none fully automate the cross-stack DSE.
Accelergy~\cite{accelergy}, SNSv2~\cite{snsv2}, and TNNGen~\cite{tnngen} offer no automation at all.
Aladdin~\cite{aladdin}, DSAGEN~\cite{dsagen}, and OverGen~\cite{overgen} automate part of the space, holding the software fixed and sweeping only the hardware.
Exploring alternative software thus means re-implementing it for each design point.
Charm~\cite{charm} alone includes full automation, but only over analytical architecture models.

Beyond automated DSE, most frameworks demand high cross-stack expertise.
Aladdin requires the designer to be proficient in parallel C programming~\cite{alladin}.
A designer without extensive parallel-programming knowledge will make naive design choices, such as generating non-vectorized hardware.
Similarly, DSAGEN~\cite{dsagen} and OverGen~\cite{overgen} require the designer to express parallelism in C with compiler directives, and SNSv2~\cite{snsv2} requires RTL expertise.

No single framework supplies the programmability to automate cross-stack DSE, motivating a framework that automates the exploration of emerging systems.

\subsection{Explainability}
Understanding how a design decision drives performance and cost requires visibility of finer-grain scope within the system.
When systems are reduced to a single end-to-end result, designers are left guessing the hierarchical behavior.
Using the same GEMM example from Section~\ref{subsec:programmability}, mapping tiles to the array, and the computation itself, carry separate costs that aggregate into the end-to-end total.
To isolate one of these events, a designer must define the scope of the metric and report it at that granularity, which existing frameworks do not allow.
Moreover, if frameworks do not expose how these metrics are aggregated, designers are left unable to trace a cost back to the interactions that produced it.

Most existing frameworks report only a single end-to-end number, obscuring fine-grain system behavior.
Charm~\cite{charm}, SNSv2~\cite{snsv2}, DSAGEN~\cite{dsagen}, OverGen~\cite{overgen}, and TNNGen~\cite{tnngen} report only a single end-to-end result.
Accelergy~\cite{accelergy} and Aladdin~\cite{alladin} report at a finer per-component scope, yet neither lets the user define it nor reveals how each number is aggregated.
EDA tools give hierarchical PPA, but that hierarchy stops at the circuit level.
Recent EDA works have begun to pursue this explainability~\cite{gem5_copilot, oac_nsf, archsim_harmful, explainable_dse}, but none offers a comprehensive cross-stack view.

This lack of explainability leaves designers of emerging systems unable to attribute cost, motivating a framework that exposes these fine-grained interactions.
\section{\name Representation}

\begin{figure}[!t]
  \centering
  \includegraphics{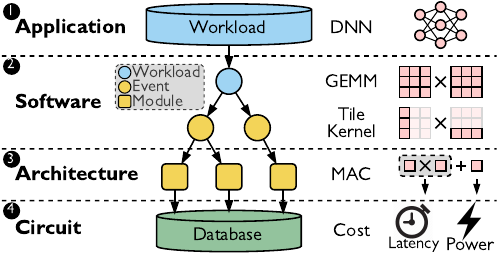}
  \caption{Overview of \name.}
  \label{fig:a-graph_flow}
\end{figure}

To address flexibility, we introduce Architecture-Graph (\name), a unified representation that captures interactions across the system stack.
Figure~\ref{fig:a-graph_flow} illustrates \name, representing a user-defined system across the \textit{application}, \textit{software}, \textit{architecture}, and \textit{circuit} stacks.
Moreover, \name separates cost aggregation from performance models, enabling tightly coupled cost co-simulation.
Finally, \name is designed to accept user-defined metrics that are relevant to the system being represented.
Through these, \name is flexible enough to extend to new systems as they emerge.

\label{sec:a-graph}

\subsection{Unified Representation}
\label{subsec:unified_modeling}

\name represents the full system within a single event-based graph, unifying the application, software, architecture, and circuit stacks.
Each stack is expressed through events and their decomposition, i.e., how they break down into sub-events, capturing interactions across layers within the same graph.
Representing the stack and its interactions hierarchically, \name lets designers fully describe the target system without being constrained to a particular stack.
This allows \name to separate the system representation from the cost modeling that drives it, resulting in a tightly coupled cost co-simulation.

In Figure~\ref{fig:a-graph_flow}, \name is organized into four system stacks: \textit{application}, \textit{software}, \textit{architecture}, and \textit{circuit}.

\minisection{\circled{1} Application:}
At the top of the stack, the application layer represents a user-defined specification of the high-level program or algorithm that serves as the input workload to \name, highlighted as the \textit{blue} root event.
By leaving the application layer user-defined, \name extends easily to emerging applications without requiring a fixed compiler flow or pre-existing software stack.

As an example (Figure~\ref{fig:a-graph_flow}), consider a deep neural network (DNN) model as the application.
This DNN workload has parameters (e.g., batch size, layer count, hidden dimension) that describe the workload and drive the decomposition of the application into software events.

\minisection{\circled{2} Software:}
The software layer builds out the middle of the graph, decomposing the workload into architecture events and driving the hierarchical mapping within \name.
For each software event, designers provide performance models that generate weighted edges to sub-events, quantitatively encoding this decomposition.
Each performance model is isolated to the specific software event it is designed for.
This runs performance models locally, effectively separating them from cost aggregation.
Moreover, the event granularity is user-defined, supporting the trade-off between evaluation speed and fidelity.

Building on the DNN example, the performance models decompose the DNN into GEMM layers, then further into tile sizes that match the compute array.


\begin{figure}[!t]
  \centering
  \includegraphics{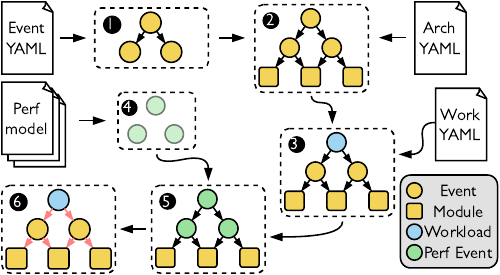}
  \caption{\name construction.}
  \label{fig:agraph-generation}
\end{figure}

\minisection{\circled{3} Architecture:}
\label{subsubsec:graph_construction_architecture}
The architecture layer defines the micro-architecture blocks that compose the system design.
These architecture modules (e.g., compute units, memory blocks, gates) are instantiated as sub-events of the software layer's leaf nodes.
The architecture layer is a flattened description that makes no assumptions about structural hierarchy, instance relationships, or spatial dependencies.
This abstraction keeps module definitions fully user-configurable, letting designers adjust granularity and representation to their modeling needs.

Continuing the DNN example, the matrix tiles from the software layer map to the compute array and break down into the multiply-accumulate (MAC) operations and other hardware events (e.g., memory, registers, FIFOs) each tile requires.

\minisection{\circled{4} Circuit:}
\label{subsubsec:graph_construction_circuit}
At the bottom of the stack, the circuit layer connects architecture modules to technology-specific metrics.
Each module references a database of precomputed circuit-level results (e.g., from EDA tools), or outputs from cost simulators (e.g., CACTI7~\cite{cacti7}), enabling rapid retrieval of accurate metrics without performing full system simulation.
By allowing users to define where metrics are sourced, e.g., CMOS and $\text{mm}^2$, FPGAs and LUTs, superconducting and JJs, \name can easily extend to emerging systems without requiring mature EDA tools or pre-existing cost models.

Each architecture module (e.g., a MAC unit) carries a hardware cost (e.g., area, power) that supplies \name with circuit-level metrics for aggregation.
This decomposition produces a graph that represents a unified system across all stacks.

\subsection{Graph Construction.}
\label{subsec:graph_construction}

Figure~\ref{fig:agraph-generation} depicts how \name is constructed.
\name starts as a directed acyclic graph (DAG), where nodes represent system events and edges point from events to their sub-events.
Each non-leaf event is supplied with a performance model that defines how it decomposes into sub-events.

Besides the circuit stack, which is either represented by a database of precomputed costs or retrieved from a plugged-in simulator, each stack is defined with a YAML file.
In Figure~\ref{fig:agraph-generation}, \circled{1} the graph starts with the event YAML, or software stack, which defines the graph structure of the system.
\circled{2}-\circled{3} add the application workload to the root and architecture modules to the leaves of the graph.
\circled{4}-\circled{5} map the performance models, which detail how events decompose into sub-events, onto the event graph.
Finally, \circled{6} applies the performance models, supplying each edge with a weight.

This transforms \name from a DAG to a weighted directed acyclic graph (WDAG).
The resulting WDAG captures the user-defined behavior across every stack and is ready to report cost across the system.

\begin{figure}[!t]
    \centering
    \includegraphics{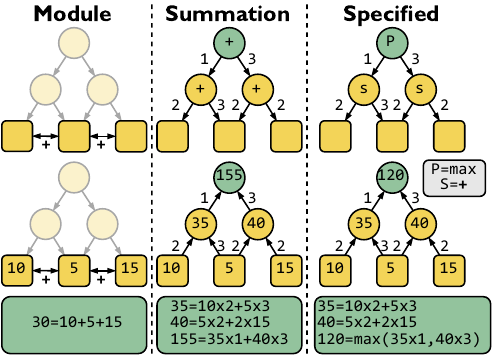}
    \caption{\name metric aggregation. Circles represent events, while squares represent modules.
             The top graph illustrates each aggregation functionality, while the bottom graph provides the aggregation flow and final value, shown in green. The green boxes display equivalent equations.
             }
    \label{fig:aggregation}
\end{figure}

\subsection{Metric Aggregation}
\label{subsec:metric_aggregation}

\name supports customized metric aggregation, letting designers define metrics for the target system and how those metrics aggregate into cost.
This customization allows \name to extend to emerging systems whose cost is defined by unique metrics.

Metric aggregation proceeds by traversing the graph in topological order, aggregating along weighted edges from module nodes up the graph.
Weighted edges are treated multiplicatively, scaling metrics according to the number of times a child event is invoked by its parent.
This process produces system-level cost by aggregating values from user-defined metrics (e.g., area, power, latency), exposing how the system's cost breaks down.

To give designers cost beyond end-to-end results, \name provides scope-based metric retrieval.
Designers select the scope at which metric retrieval occurs, aggregating only a subgraph within the system graph.
This subgraph can be constructed by selecting any application or software event as its scope.
As an example, let us revisit the DNN system from Figure~\ref{fig:a-graph_flow}.
If a designer wanted to find the cost of a tile mapping instead of the entire DNN workload, then a tile event could be selected as the scope.
The resulting cost would only consider this tile event and those below it, giving insight into costs that build the end-to-end results.

\name currently supports three types of aggregation, visualized in Figure~\ref{fig:aggregation}.
\begin{itemize}
    \item \textbf{\textit{Module}}: Directly aggregates metrics of architecture modules, independent of application and software influences (e.g., area, leakage power).
    \item \textbf{\textit{Summation}}: Accumulates metrics across edges, collapsing to a single value at the root node.
    This applies to aggregation with no temporal dependencies, such as dynamic energy.
    \item \textbf{\textit{Specified}}: Supports two interaction modes, \textit{sequential} and \textit{parallel}.
    Sequential events, similar to summation, accumulate incoming metrics, while parallel events take the maximum, modeling interactions like runtime where parallel execution is bounded by the slowest event.
    Dependency-induced stalls and pipelining are not captured by aggregation and must come from the user's performance model or an external trace.
\end{itemize}

\name's metric aggregation is as flexible as a spreadsheet~\cite{grossman2007spreadsheet, timsit2010using, yoder2005architectural}, but tailored to cross-stack cost modeling for user-defined systems.
Therefore, we coin it as \emph{a topological spreadsheet}.
By aggregating cost at user-defined scopes, \name is flexible and explainable.

\section{\thiswork Framework}
\label{sec:architecture}
\begin{figure*}[!t]
  \centering
  \includegraphics{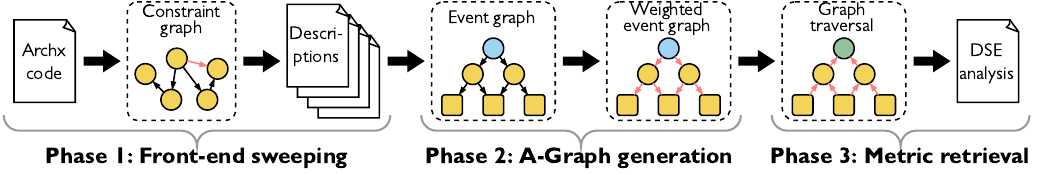}
  \caption{Overview of \thiswork.
  }
  \label{fig:archx_overview}
\end{figure*}

To further address programmability and explainability, we develop \thiswork to implement \name at its core for fast DSE, with the workflow shown in Figure~\ref{fig:archx_overview}.
In Phase 1, \thiswork provides a user-friendly programming interface to automatically generate and sweep design points under user-defined constraints.
Then, phase 2 constructs an \name for each configuration.
Finally, Phase 3 incorporates scope-based metric retrieval and aggregation exposure to analyze and understand each design point at user-defined granularity.

\subsection{Phase 1: Front-end Sweeping}
\label{subsec:frontend}
\thiswork supports a user-friendly front-end interface to define and constrain systems.
The system description is then automatically swept across valid design points within the provided constraints, generating description files.
Moreover, \thiswork does not require developers to have cross-stack expertise.
For example, Aladdin~\cite{alladin} mandates optimal C code for processor design.
The following sections detail the process of defining, constraining, and generating the design space in \thiswork.

\begin{lstlisting}[float=!t, language=Python, caption={MAC array front-end code.}, label={lst:mac_frontend}]
def description(path):
  agraph = AGraph(path=path)
  arch, event, mtrc, wkld = agraph.desc

  # Add workloads (Application)
  W = wkld.add(['GEMM'])
  batch = W.add(name='batch', value=4)
  dim = W.add(name='dim', value=[32, 64],
              sweep=True)
  
  # Add events (Software)
  event.add(name='GEMM', subevent=['tile'])
  event.add('tile', ['sram', 'mac', 'p-reg'])
  
  # Add modules and attributes (Architecture)
  arch.add_attr(tech='7nm', interface='cmos')
  arr_inst = [[x*2, x*2] for x in range(2, 6)]
  sram_width = [[x*2] for x in range(2, 6)]

  mac=arch.add(name='mac', inst=arr_inst,
    tag=['pe'], query={'class':'mac'})
  pReg=arch.add(name='p-reg', inst=arr_inst,
    tag=['pe'], query={'class':'reg', 'width': 16})
  sram=arch.add(name='sram', inst=[1], tag=['mem'],
    query={'class':'sram', 'interface':'cacti7',
           'depth':128, 'width':sram_width})
  
  # Add metrics (Circuit)
  mtrc.add(name='area', unit='mm^2', aggr='module')
  mtrc.add(name='energy', unit='nJ', aggr='sum')
  mtrc.add(name='runtime', unit='ms', aggr='spec')
  
  # Add sweeping constraints (Constraint)
  agraph.injection([pReg['inst'], mac['inst']])
  agraph.exclusion(a=mac['inst'], b=dim,
                   cond=lambda a, b: a[0] == b)
  agraph.condition(a=mac['inst'], b=sram['width'],
                   cond=lambda a, b: a[0] == b)
  return agraph.generate()
\end{lstlisting}

Listing~\ref{lst:mac_frontend} presents the programming interface with a similar example as Figure~\ref{fig:a-graph_flow}, using a MAC array (MA) running a tiled GEMM workload.
At the start of the listing (lines 1-3), an \name object and its descriptions are instantiated.

\minisection{Workload Description:}
\label{subsubsec:workload_description}
The workload description (lines 5-9) defines a workload (i.e., \textit{GEMM}) and its parameters (i.e., batch size (\textit{batch}) and matrix dimension (\textit{dim})).

\thiswork provides two methods for defining workload parameters: static and sweeping.
Static parameters use a value as-is. 
Here, the batch is static, only considering a single batch size.
Sweeping parameters iterate over the provided list.
The dim parameter is a sweeping parameter, which will generate configurations for each element within the list.
To distinguish a sweeping parameter from a static one, an additional sweep argument is needed, as seen on line 9.

\minisection{Event Description:}
\label{subsubsec:event_description}
Each workload must then be defined as a root node in the software description (lines 11-13), creating an event with the same name as the workload.
Events propagate from root to leaf events by recursively invoking sub-events until reaching the architecture modules.
The root \textit{GEMM} then defines \textit{tile} as its sub-event, which further decomposes into the architecture modules \textit{mac}, \textit{p-reg}, and \textit{sram}.
Each event is associated with a performance model that specifies such decomposition and computes the edge weights to each sub-event, e.g., how many \textit{tiles} under \textit{GEMM}.

\minisection{Architecture Description:}
\label{subsubsec:architecture_description}
The architecture description instantiates modules and defines their attributes (lines 15-26).
First, global attributes are default for all architecture modules, and can be overwritten by local attributes within each module.
This is highlighted in the \textit{sram} module (line 24), which uses the \textit{cacti7} interface as its cost database instead of the global \textit{cmos} interface for compute logic.

Then, each module is instantiated with three key properties: \emph{instance}, \emph{tag}, and \emph{query}.
The \emph{instance} property specifies the number of module instances, which can also be static (a list) or sweeping (a list of lists), similar to workload parameters.
Next, the \emph{tag} property groups modules for scoped metric retrieval (Section~\ref{subsec:metric_retrieval}).
Finally, the \emph{query} property defines the configuration to retrieve metrics from a module database, specified by class.
Each query is user-defined by the database, consistent with \name's flexibility for defining new metrics and cost databases.

\minisection{Metric Description:}
\label{subsubsec:metric_description}
The metric description lists the available metrics and specifies how they are aggregated when traversing the graph for metric retrieval.
Each metric is defined with a name, unit, and aggregation type, e.g.,  \textit{area}, \textit{energy}, and \textit{runtime} (Section~\ref{subsec:metric_aggregation}).

\minisection{Constraint Graph:}
By default, \thiswork sweeps all parameters, which generates invalid design points.
To constrain the design space to valid parameter combinations, \thiswork provides design constraining to remove invalid or unwanted configurations.
This sweeping is implemented with three types of constraints: \emph{Injection}, \emph{Exclusion}, and \emph{Condition}.
\begin{itemize}
    \item 
    \emph{Injection} constraints enforce a one-to-one mapping between parameters, such as aligning the MA array's processing elements and pipeline registers.
    \item 
    \emph{Exclusion} constraints remove invalid parameter combinations, such as removing MA array sizes that are larger than the matrix dimensions.
    \item  
    \emph{Condition} constraints enforce a mapping between parameters that meet the condition, while removing combinations that do not, such as ensuring the MA array size aligns with the SRAM width.
\end{itemize}

\begin{figure*}[!t]
  \centering
  \includegraphics{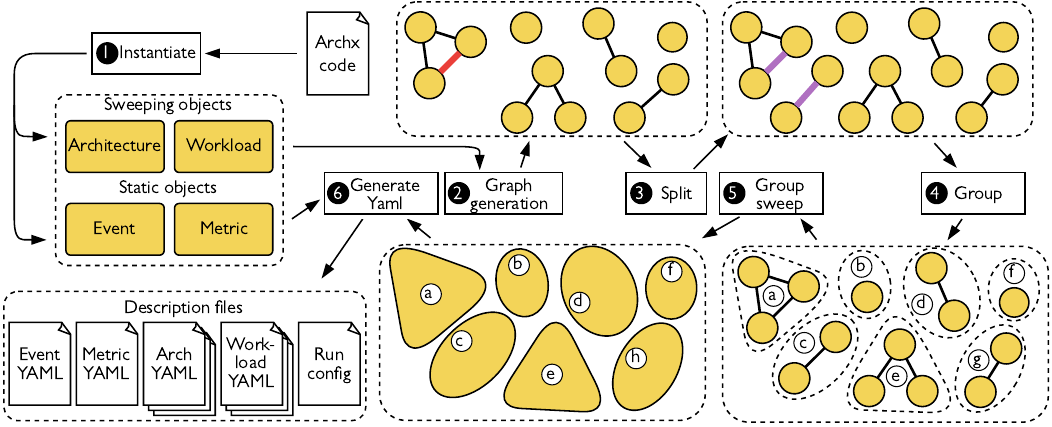}
  \caption{\thiswork constraint graph and description generation. 
  Nodes represent workload or module parameters, while edges denote constraints. 
  Red edges indicate condition constraints, while purple edges are the result of splitting these conditions.}
  \label{fig:constraint_graph_sweeping}
\end{figure*}

Figure~\ref{fig:constraint_graph_sweeping} outlines how designs are constrained in \thiswork.
It represents the design space as a constraint graph constructed from the system description and user-defined constraints.
\circled{1} Static objects, such as event and metric descriptions, are directly forwarded to generate YAML description files.
Sweeping objects, including architecture and workload descriptions, are used to construct the initial constraint graph in \circled{2}.
\circled{3} All \textit{condition} constraints are split into multiple non-condition constraints, to define the valid and invalid combinations from the conditions.
\circled{4} Nodes connected by edges are grouped into subgraphs, and each subgraph is reduced to produce local sweeps.
\circled{5} A global sweep across groups then considers all possible combinations of these subgraphs.
\circled{6} Finally, the swept architecture and workload combinations are joined with static objects to generate the YAML files.
This produces description files that sweep the design space, constraining configurations to only valid design points.

Through this front-end interface, \thiswork empowers designers to define \name descriptions and constraints, automatically generating and sweeping the design space at all specified design points, demonstrating \thiswork’s high programmability.

\begin{lstlisting}[float=!t, language=Python, caption={Performance model code.}, label={lst:mac_perf}]
def GEMM(arch_dict, wkld_dict):
  # Retrieve workload and architecture parameters
  mac_dim = arch_dict['mac']['inst'][0]
  M, K, N = wkld_dict['config']['dim'] #matrix dims
  m, k, n = mac_dim[0]    # tile dims
  cnt = (M/m) * (K/k) * (N/n) # Array mappings
  tile_dict={'count': cnt, 'aggr':'sequential',
             'factor':{'runtime':2}}  
  # Add to performance dict
  perf_dict = {'subevent': {'tile': tile_dict}}

  return perf_dict

def tile(arch_dict, wkld_dict):
  # Retrieve architecture parameters
  perf_dict = {}
  mac_inst = arch_dict['mac']['inst']
  reg_inst = arch_dict['p-reg']['inst']
  bitwidth = arch_dict['p-reg']['query']['width']
  swidth = arch_dict['sram']['query']['width']
  freq = arch_dict['mac']['freq']
  
  # Calculate runtime and sram events
  sram_events = (mac_inst[0]**3 * bitwidth)/swidth
  perf_dict['runtime'}={'value':1/freq,'unit':'ms'}
  # Add to performance dict
  subevents = {{'mac':   {'count': mac_inst}},
               {'p-reg': {'count': reg_inst}},
               {'sram':  {'count': sram_events}}}
               
  perf_dict['subevent'] = subevents
  return perf_dict
\end{lstlisting}

\subsection{Phase 2: \name Generation}
With the system design constrained and generated, each configuration is then used to construct an \name, following the process outlined in Section~\ref{subsec:graph_construction}.

\minisection{Performance Model:}
Every event in \thiswork requires a performance model to define the decomposition to its sub-events.
Listing~\ref{lst:mac_perf} details two performance models: the workload root event \textit{GEMM} (lines 1-12) and the leaf event \textit{tile} (lines 13-32).

The \textit{GEMM} model determines the \textit{count} of tile mappings, which is assigned to the edge from \textit{GEMM} to \textit{tile}.
Relevant parameters are first retrieved from the architecture and workload descriptions, each represented as a dictionary.
Here, the matrix and array dimensions are used to compute the total number of tile mappings to the MAC array.
The computed mappings are added to the performance dictionary, then assigned to the event edge.
The performance model specifies the \textit{aggregation} for metrics along each edge.
Two aggregation types are available, \textit{sequential} and \textit{parallel}, which outline the two options for aggregating specified metrics as detailed in section~\ref{subsec:metric_aggregation}.
Given that our design includes only a single MAC array, one mapping is computed at a time.
The runtime of each mapping must wait for the previous to finish, therefore the runtimes accumulate, corresponding to sequential aggregation.
With multiple MAC arrays, aggregation could instead be parallel.

Additionally, each edge can have an associated factor, which applies multiplicative scaling of the event count.
This is commonly used to parameterize the design or account for hardware utilization.
Here, a factor of 2 is applied to runtime to model a non-pipelined MA that requires two cycles.

Unlike the workload model, which focuses on tiling, the leaf event \textit{tile} model specifies computations for its metrics (e.g., runtime in line 25), in addition to setting the event count for each module along the edge.
In this example, runtime is calculated from the MAC frequency, while the counts for the \textit{mac} and \textit{p-reg} modules are determined from the array shape.
Next, the \textit{sram} event is computed by scaling the number of reads and writes by the array shape and bitwidth, so the performance graph captures both computation and memory cost for the MA operation.

The resulting graph after applying each performance model is a weighted \name that captures the hierarchical interactions between each stack within the described system.

\subsection{Phase 3: Metric Retrieval}
\label{subsec:metric_retrieval}
\thiswork empowers designers with detailed metric analysis through its scope-based retrieval and aggregation exposure.

\minisection{Scope-based Metric Retrieval:}
In order to fully understand a system's behavior, designers must be able to view resulting metrics at varying levels of granularity within the system.
To enable this, \thiswork provides scope-based metric retrieval, allowing designers to select a metric and define the scope of interest for analysis.
There are four knobs for such retrieval:
\begin{itemize}
    \item Event: limits scope to an event and its sub-events.
    \item Tag: limits scope to events that decompose into modules sharing the same tag.
    \item Workload: Selects an event that is treated as a root node, aggregating the sub-tree of events and modules under it.
    \item Module: limits scope to a single module, aggregating all events that decompose into it.
\end{itemize}
The event and tag knobs are mutually exclusive, but either can combine with the workload knob.
Graph traversal always starts from the leaf module nodes, aggregating upward until reaching the root of the specified scope, allowing designers to focus on the scope of interest for hierarchical analysis.

\begin{figure}[!t]
  \centering
  \includegraphics{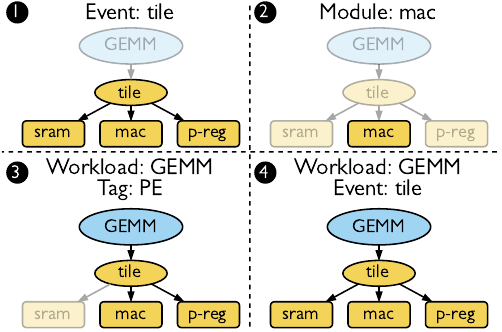}
  \caption{Scope-based metric retrieval in \thiswork.}
  \label{fig:archx_query_scope}
\end{figure}

Figure~\ref{fig:archx_query_scope} showcases several examples of metric retrieval and scope selection.
\circled{1} With no workload specified, the metrics of a single \textit{tile} event will be retrieved from all modules.
\circled{2} retrieves the metrics of the specific module \textit{mac}, while all others are ignored.
\circled{3}-\circled{4} both specify \textit{GEMM} as the workload.
\circled{3} retrieves the metrics based on the tag \textit{pe}. 
Therefore, \textit{sram} metrics, with no tag of \textit{pe}, are excluded, as detailed in Listing~\ref{lst:mac_frontend} line 24.
\circled{4} With a workload specified, the metrics of all \textit{tile} events will be retrieved, instead of a single \textit{tile} in \circled{1}.

\minisection{Aggregation Exposure:}
In addition to scope-based retrieval, \thiswork exposes the aggregation of metrics along each edge in the graph.
In \thiswork, as events are aggregated during metric retrieval, the hierarchical aggregation is saved as log files or directly output to the terminal.
This allows designers to track how metrics are aggregated for better understanding.

Through scope-based metric retrieval and aggregation exposure, \thiswork enables designers to understand and analyze all design points at desired granularity with high explainability.
\section{Experimental Setup}
\label{sec:experimental_setup}

\begin{table}[!t]
    \centering
    \caption{Implemented case studies.}
    \begin{adjustbox}{max width=\linewidth}
    \begin{tabular}{ccccc}
        \toprule
        \textbf{System}              & \textbf{Technology}              & \textbf{Node}              & \textbf{Application} & \textbf{Architecture}                \\
        \midrule
        \multirow{3}{*}{\emph{Emerging}}    & \multirow{2}{*}{Superconducting} & \multirow{2}{*}{10 KA/cm2} & FIR                  & Tap array~\cite{superconducting_fir} \\
        \cmidrule{4-5}
                                     &                                  &                            & CNN                  & PE array~\cite{superconducting_cnn}  \\
        \cmidrule{2-5}
                                     & CMOS                             & 7nm                        & TNN                  & TNN column~\cite{tnngen}             \\
        \midrule
        \multirow{4}{*}{Conventional} & \multirow{4}{*}{CMOS}            & \multirow{2}{*}{45nm}      & FFT                  & Butterfly array                      \\
        \cmidrule{4-5}
                                     &                                  &                            & GEMM                 & Systolic array                       \\
        \cmidrule{3-5}
                                     &                                  & 45nm                       & GEMM                 & RISC-V core~\cite{riscv}             \\
        \cmidrule{3-5}
                                     &                                  & 4nm                        & LLM                  & NVIDIA H200~\cite{nvidia_h200_spec}  \\
        \bottomrule
    \end{tabular}
    \end{adjustbox}
    \label{tab:case_studies}
\end{table}

To validate \name and \thiswork's usability, we implement multiple case studies on \thiswork, for both emerging and conventional systems (Table~\ref{tab:case_studies}). We compare simulation fidelity with that of EDA tools for CMOS technologies, and against prior works for non-CMOS technologies.

\subsection{Emerging Systems}
\minisection{Superconducting:}
Superconductors are metals that exhibit zero electrical resistance at low temperatures, e.g., 10mK$\sim$70mK, promising ultra-low power and ultra-high efficiency~\cite{a_case_for_superconducting}.
A popular superconducting logic family is single flux quantum (SFQ) logic, which uses the Josephson Junction (JJ) as its fundamental switching element~\cite{a_case_for_superconducting, RSFQ_logic}.
Superconducting circuits transmit information by quantizing magnetic flux, encoding data as discrete flux quanta (spikes) rather than continuous voltage levels~\cite{a_case_for_superconducting}, achieving a substantial decrease in power compared to conventional CMOS designs~\cite{superconducting_energy, rql_energy} while unlocking higher clock frequencies~\cite{RSFQ_tff, RSFQ_logic}.

We validate superconducting through a finite impulse response (FIR) tap array~\cite{superconducting_fir} and a convolutional neural network (CNN) PE array~\cite{superconducting_cnn}. 
Both architectures are implemented using the open-source MIT-LL SFQ5ee 10 kA/cm$^2$ process~\cite{MITLLSFQ} and simulated with WRSPICE. 
The resulting metrics are area as the number of Josephson Junctions (JJs), dynamic and leakage power, and throughput.

\minisection{Neuromorphic:}
Neuromorphic computing is an emerging AI workload that models the behavior of biological neurons, commonly implemented as either temporal neural networks (TNNs)~\cite{neocortical_computation, tnnarch} or spiking neural networks (SNNs)~\cite{liu2015spiking, 10.1145/3316781.3317870, adam20163, ankit2017resparc}.
TNNs operate on temporal-coded spikes to reduce switching activity, while SNNs rely on rate-coded spikes.

In this study we focus on TNNs, constructing a single TNN column, a layer of a TNN.
We validate results against the \textit{full EDA flow} by using Cadence Genus 21.19 for synthesis and Cadence Innovus 19.10 for place-and-route with the ASAP7~\cite{ASAP7} and customized TNN7~\cite{nair2022tnn7} process nodes and PDKs.
The hardware is implemented as either standard cells in Verilog, or circuit macros, with floorplan utilization set to 70\% for fair comparison.
The TNN utilizes a faster 100 MHz clock, and a slower gamma clock of 0.378 MHz.

\subsection{Conventional CMOS Systems}
We model conventional CMOS systems, showcasing the flexibility of \name.
All hardware is implemented similarly to the neuromorphic methodology, using NanGate45~\cite{NanGate45} as the process node.
The hardware for the fast Fourier transform (FFT), GEMM, and RISC-V is designed at a 400 MHz clock frequency.
Additionally, we scale evaluation to the system level by deploying a large language model (LLM), GPT2~\cite{radford2019language}, on an NVIDIA H200 NVL system to compare against \thiswork.

\section{Case Study}
\label{sec:case_study}

\subsection{Simulation speedup}
\label{subsec:eda_speedup}
\begin{figure}[!t]
  \centering
  \includegraphics{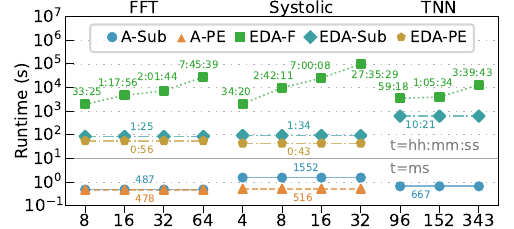}
  \caption{EDA vs \thiswork (A) runtime across configurations. 
  -PE and -Sub denote running \name or EDA flow for the full PE and submodules.
  -F denotes running full design EDA flow.
  }
  
  \label{fig:runtime}
\end{figure}

Figure~\ref{fig:runtime} compares the runtime of the full EDA flow, module database generation, and \thiswork for CMOS. Compared to the full EDA flow, \thiswork is up to $10^5\times$ faster, or $10^3\times$ faster including database generation.
Moreover, \thiswork generates all configurations from a single description within milliseconds.
With the short generation and execution times, Archx is capable of exploring thousands of design points in the same time it takes EDA tools to retrieve results for a single configuration.
\emph{Both demonstrate \thiswork's programmability, delivering fast DSE compared to baseline results.}

\begin{figure}[!t]
  \centering
  \includegraphics{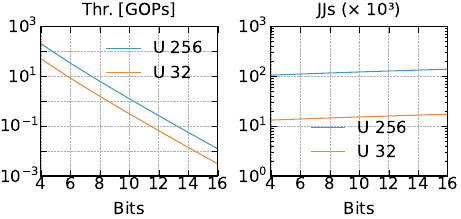}
  \caption{Validation of FIR bitwidth (x axis).
  }
  \label{fig:fir_study}
\end{figure}

\subsection{Emerging Systems}
\subsubsection{Superconducting}
This section covers both FIR and CNN case studies.

\minisection{Finite Impulse Response:}
An FIR filter computes a finite-duration impulse response using sequential taps, each comprising of a shift register, multiplier, and adder.
The multiplier and adder are implemented with a U-SFQ data processing unit (DPU).
Leveraging superconducting and unary computing, an FIR can be implemented with improved efficiency at low bit resolutions~\cite{superconducting_fir}.


Without access to the original throughput and area values, we reproduce their reported trends in Figure~\ref{fig:fir_study}.
Our results yield nearly identical lines for the U-SFQ 32- and 256-tap FIR arrays (U 32, U 256) from~\cite{superconducting_fir}.
We implement the DPU (U-SFQ multiplier and adder) in \thiswork and model both dynamic and leakage power, as shown in Table~\ref{tab:fir_power}.
Our results exhibit low relative error for both.
Given that leakage power dominates total power in superconducting designs, the absolute error in total power is also small, yielding nearly identical results to the baseline.
\emph{\thiswork's ability to achieve high fidelity stems from its flexibility, enabling modeling for emerging systems.}

\begin{table}[!t]
    \centering
    \caption{Validation of FIR power results.}
    \begin{adjustbox}{max width=\linewidth}
    \begin{tabular}{cccc}
        \toprule
        \textbf{Metric} & \textbf{Archx} & \textbf{Baseline} & \textbf{Relative error (\%)}\\
        \cmidrule{1-4}
        \textbf{Dynamic power ($\mu$W)} & 8.125 & 8.4 & -3.27 \\
        \cmidrule{1-4}
        \textbf{Leakage power (mW)} & 8.4 & 8.4 & 1.81$\times10^{-14}$ \\
        \bottomrule
    \end{tabular}
    \end{adjustbox}
    \label{tab:fir_power}
\end{table}

\minisection{Convolutional Neural Network:}
CNNs are deep learning models that excel at feature extraction, specifically for image classification.
Similar to GEMMs, these networks require highly parallelized hardware for efficient computation.
Prior superconducting work has proposed a 3D PE array, with a scale-up array size of $64\times64\times32$~\cite{superconducting_cnn}.

We implement this 3D PE array in \thiswork to study a matrix convolution workload, reporting area, power, and throughput, as shown in Table~\ref{tab:sc_cnn}.
Our results match the baseline across all metrics other than area.
In place of JTL chains, common for long connections, the baseline utilizes passive transmission lines (PTLs) as overhead.
Including this overhead reduces the error from 15\% to 3\%.
Table~\ref{tab:sc_cnn_breakdown} further breaks down resources that are not reported in~\cite{superconducting_cnn}.
\emph{Both of these demonstrate \thiswork's explainability, exposing the system behavior that drives the understanding and attribution of emerging systems.}

\begin{table}[!t]
    \centering
    \caption{Validation of CNN Results.
    3D PE array has a dimension of $64\times64\times32$.}
    \begin{adjustbox}{max width=\linewidth}
    \begin{tabular}{cccc}
        \toprule
        \textbf{Metric} & \textbf{Archx} & \textbf{Baseline} & \textbf{Relative error (\%)}\\
        \cmidrule{1-4}
        \textbf{Area w/o overhead (JJ)} & 13.1M & 15.4M & -14.89 \\
        \cmidrule{1-4}
        \textbf{Area w/ overhead (JJ)} & 14.9M & 15.4M & -2.97 \\
        \cmidrule{1-4}
        \textbf{Power (W)} & 3.8 & 3.9 & -2.54  \\
        \cmidrule{1-4}
        \textbf{Throughput (TMACs)} & 409.6 & 409 & 0.15  \\
        \bottomrule
    \end{tabular}
    \end{adjustbox}
    \label{tab:sc_cnn}
\end{table}

\begin{table}[!t]
    \centering
    \caption{Validation of CNN array breakdown. 
    S denotes splitters.
    NDRO is for Non-Destructive Readout memory.}
    \begin{adjustbox}{max width=\linewidth}
    \begin{tabular}{cccccc}
        \toprule
        \textbf{Module} & \textbf{MAC} & \textbf{MAC S} & \textbf{Weight S} & \textbf{Input S} & \textbf{NDRO} \\
        \cmidrule{1-6}
        \textbf{Area (JJ)} & 12.3M & 393.2K & 393.2K & 393.2K & 448 \\
        \cmidrule{1-6}
        \textbf{Power (mW)} & 3801.08 & 785.86 & 785.86 & 785.86 & 0.6 \\
        \bottomrule
    \end{tabular}
    \end{adjustbox}
    \label{tab:sc_cnn_breakdown}
\end{table}

\subsubsection{Neuromorphic Computing}
\label{subsec:neuromorphic_case_study}
\begin{figure}[!t]
  \centering
  \includegraphics{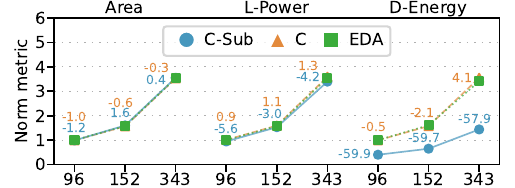}
  \caption{Validation of TNN input-size (x axis). 
  Each TNN column (C) has 96, 152, or 343 input neurons, with fixed 2 output neurons.
  All metrics are normalized to the full EDA results of the 96-input column.
  C-Sub denotes the fine-grained \thiswork implementation using submodules of the column.
  }
  
  \label{fig:tnn_array}
\end{figure}

Figure~\ref{fig:tnn_array} illustrates \thiswork's modeling of a TNN column architecture across 3 different input neuron sizes.
Both area and leakage power show low relative error, capped at 1.3\% for both C-Sub and C configurations.
The discrepancy in C-Sub's dynamic energy arises from submodule EDA macro results not accounting for the actual circuit fanout load.
When capturing this load in the column correctly, the gap is closed, as seen in C.
\textit{These results underscore the importance of selecting the correct module granularity for high fidelity, while trends still remain present at lower fidelity, demonstrating the flexibility for apples-to-apples comparison.}

\begin{figure}[!t]
  \centering
  \includegraphics{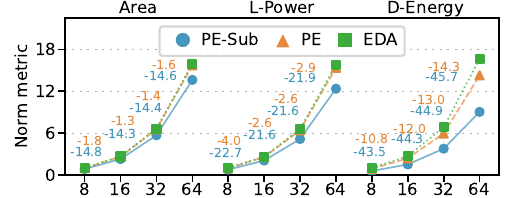}
  \caption{Validation of FFT array size (x axis). 
    All metrics, area, leakage power (L-Power), and dynamic energy (D-Energy), are normalized to the full EDA results of the $8\times3$ array.
    Relative errors (\%) are shown above each plot.
    PE in orange means \thiswork uses the full butterfly unit directly in the database.
    PE-Sub means \thiswork's database uses submodules of a PE.
    }
    
  \label{fig:fft_array}
\end{figure}

\subsection{Conventional CMOS Systems}
\subsubsection{Fast Fourier Transform}
\label{subsubsec:FFT}
We implement a flattened FFT with all butterfly stages fully pipelined.
With $2N$ inputs, the FFT array has a shape of $N\times log_2N$.
This flattened FFT allows one-time input mapping without involving memory accesses.

\begin{figure}[!t]
  \centering
  \includegraphics{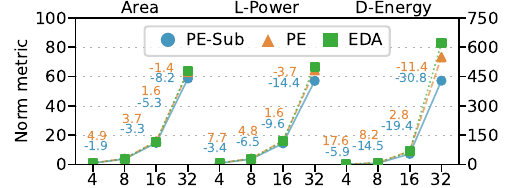}
  \caption{Validation of systolic array size (x axis).
  All metrics are normalized to the full EDA results of the $4\times4$ array.
  Notations follow these in Figure~\ref{fig:fft_array}. 
  }
  
  \label{fig:sys_array}
\end{figure}

\begin{table}[!t]
    \centering
    \caption{Systolic array configuration and full \thiswork results.
    Array dimension is the shape of the systolic array.
    Matrix dimension represents m$\times$k$\times$n of GEMM.}
    \begin{adjustbox}{max width=\linewidth}
    \begin{tabular}{ccccc}
        \toprule
        \textbf{Array dim.} & 4x4 & 8x8 & 16x16 & 32x32 \\
        \cmidrule{1-5}
        \textbf{Matrix dim.} & 16x16x16 & 32x32x32 & 64x64x64 & 128x128x128 \\
        \cmidrule{1-5}
        \textbf{Throughput ($GFLOPs$)} & 6.4 & 25.6 & 102.4 & 409.6 \\
         \cmidrule{1-5}
         \textbf{Compute Area ($\mu m^2$)} & 32.96 & 127.30 & 500.09 & 1982.18 \\
         \cmidrule{1-5}
         \textbf{Compute D-Energy ($nJ$)} & 16.44 & 125.00 & 973.77 & 7685.43 \\
         \cmidrule{1-5}
         \textbf{Compute L-Power ($mW$)} & 0.69 & 2.65 & 10.38 & 41.10 \\
         \cmidrule{1-5}
        \textbf{Total SRAM ($KB$)} & 4 & 16 & 64 & 256 \\
        \cmidrule{1-5}
        \textbf{SRAM Area ($\mu m^2$)} & 22.31 & 77.25 & 307.13 & 1133.88 \\
        \cmidrule{1-5}
        \textbf{SRAM D-energy ($nJ$)} & 0.60 & 3.73 & 26.60 & 191.91 \\
        \cmidrule{1-5}
        \textbf{SRAM L-Power ($mW)$} & 0.52 & 1.82 & 6.87 & 24.35 \\
        \bottomrule
    \end{tabular}
    \end{adjustbox}
    \label{tab:sys_thr}
\end{table}

Figure~\ref{fig:fft_array} shows fidelity of the FFT when scaling the array size.
We evaluate array sizes ranging from $8\times3$ to $64\times6$, with all multipliers and adders in bfloat16~\cite{bfloat16_paper}.
At the PE granularity, the error is consistently small, capped at 15\%.
In contrast, the finer PE-Sub granularity shows larger errors when the array size grows.
This results from the sub-modules not counting the wiring congestion in the full PE. This under-estimates the resulting metrics, mainly dynamic energy.
\textit{Consistent with the neuromorphic case study (Section~\ref{subsec:neuromorphic_case_study}), these results further support choosing the appropriate module granularity for high-fidelity analysis, while lower-fidelity configurations still follow the same trends.}


%

\subsubsection{Systolic Array}
We implement several weight-stationary systolic arrays, with local buffers, for GEMM workloads.
Each PE contains a multiplier, accumulator, registers and extra control logic.
Surrounding the PE array are FIFOs for synchronization, an output accumulator for tiled partial sums, and three SRAMs for input, weight, and output buffers.

Table~\ref{tab:sys_thr} reports the array configuration together with the throughput and cost of both the compute array and the SRAM.
At 100\% utilization, throughput and compute cost scale linearly with array size, since the GEMM workload grows in proportion to the number of PEs.
The resulting memory cost is queried from CACTI7~\cite{cacti7}.

Figure~\ref{fig:sys_array} shows the array scaling of systolic arrays.
At smaller array sizes, the PE implementation outperforms the PE-Sub version in fidelity.
But at larger array sizes, the PE-Sub implementation outperforms.
\emph{Such trade-offs between fidelity and array sizes suggest that users create an ensemble of module databases for different configurations to ensure fidelity, supported through \name's flexibility and programmability.
Moreover, \thiswork showcases programmability by sweeping across 4 total array and workload combinations, while more can be supported.}
Though not included here, we use \thiswork to sweep thousands of configurations for accelerating LLMs and quantum error decoding, with costs validated against place-and-route within 15\% errors.


        

\subsubsection{RISC-V Core}
The evaluated RISC-V~\cite{riscv} core is a 5-stage, in-order processor implementing the RV32I ISA. 
The design includes standard hazard detection and forwarding logic, along with a static not-taken branch predictor.
The RV32I ISA does not include multiplication, instead being implemented via bit-serial addition.
We map a 2 $\times$ 2 GEMM workload and integrate \thiswork with an external cycle-accurate instruction trace simulator, reporting 3142 cycles (including pipeline stalls).
We place-and-route each pipeline stage, extracting their energy from simulated VCD files of circuit switching activity.
Each of these stages is then instantiated in \thiswork, with the cycle trace used as the target workload.
We intentionally exclude the cache and use this instruction-level model rather than component-level (e.g., McPAT~\cite{mcpat_paper}) to demonstrate the flexibility and programmability.

Table~\ref{tab:riscv_gemm} illustrates the area, dynamic energy, and leakage power of our simulation compared to a full place-and-route implementation.
\emph{Similar to other results, \thiswork proves high fidelity for area and leakage power comparisons, both at or below 1\%, while dynamic energy stays reasonably accurate.}

\begin{table}[!t]
    \centering
    \caption{Validation of RISC-V GEMM Results.}
    \begin{adjustbox}{max width=\linewidth}
    \begin{tabular}{cccc}
        \toprule
        \textbf{Metric} & \textbf{Archx} & \textbf{Baseline} & \textbf{Relative error (\%)}\\
        \cmidrule{1-4}
        \textbf{Area ($mm^2$)} & $1.265\times10^{-2}$ & $1.271\times10^{-2}$ & -0.510\% \\
        \cmidrule{1-4}
        \textbf{Dynamic energy (nJ)} & 27.522 & 32.295 & -14.78\% \\
        \cmidrule{1-4}
        \textbf{Leakage Power (uW)} & 256.84 & 255.15 & 1.01\%  \\
        \bottomrule
    \end{tabular}
    \end{adjustbox}
    \label{tab:riscv_gemm}
\end{table}

\begin{table}[!t]
    \centering
    \caption{Validation of GPU GPT-2 Results.}
    \begin{adjustbox}{max width=\linewidth}
    \begin{tabular}{cccc}
        \toprule
        \textbf{Metric} & \textbf{Archx} & \textbf{Baseline} & \textbf{Relative error (\%)}\\
        \cmidrule{1-4}
         \textbf{Energy (J)} & 36.307 & 36.307 & 3.057\% \\
        \cmidrule{1-4}
        \textbf{Dynamic energy (nJ)} & 421.756 & 442.553 & 0.953\% \\
        \cmidrule{1-4}
        \textbf{Leakage power (uW)} & 86.087 & 86.087 & 1.00\%  \\
        \cmidrule{1-4}
    \end{tabular}
    \end{adjustbox}
    \label{tab:gpu_gpt2}
\end{table}

\subsubsection{GPU System}
Executing a workload on a GPU involves host-side orchestration, where the CPU manages data and dispatches kernels.
Kernels are hierarchically mapped to grids and thread blocks for parallel execution across streaming multiprocessors.
AI workloads further decompose computation into operators, mapping to kernels based on parallelization strategy and hardware utilization.

Table~\ref{tab:gpu_gpt2} presents results for the GPT-2 workload~\cite{radford2019language} simulated with a batch size of 32 and sequence length of 2048.
The workload is constructed hierarchically, decomposing the application into operators and kernels.
We construct the performance and cost model by integrating an in-house GPU kernel micro-benchmarking framework based on prior works~\cite{11096369, vellaisamy2026taxbreak}, leveraging pynvml~\cite{pynvml2025} for metric measurements.
\thiswork models the constructed operator–kernel dependency graph, aggregating kernel metrics to operators and the full workload.
Baseline metrics are obtained from the same framework as end-to-end measurements, providing a consistent reference for evaluation.
\emph{\thiswork shows high fidelity, producing metrics nearly identical to the baseline.
Moreover, \thiswork retrieves results in seconds, compared to the hours required for full GPU kernel profiling, reinforcing \thiswork's programmability}.
\section{Discussion}
\label{sec:discussion}

\subsection{Related Work}
\label{subsec:related_work}

\subsubsection{Architecture Modeling}
\label{subsec:charm}
Charm provides a high-level architecture language that expresses relationships between dependent architectural parameters using analytical evaluation~\cite{charm}.
It builds an internal dependency graph to solve these relationships and apply constraints and optimizations for efficient modeling.
Both Charm and \name use high-level analytical models and constrain DSE.
\name goes further by decomposing these models into real hardware metrics for cross-stack evaluation, limiting DSE to valid design points.

\subsubsection{Performance and Cost Modeling}
There exist prior works that attempt to project architecture predictions by mapping expert C code to pre-synthesized RTL blocks and memory models~\cite{alladin}.
Additional works instead leverage lookup tables to retrieve energy costs from primitive and compound compute and memory components~\cite{accelergy}.
Finally, recent works extend graph neural networks to HDL compilation, generating a hierarchical graph representation for accurate cost estimation~\cite{snsv2}.
While all are effective, each restricts itself to a specific target or stack, limiting generality to emerging systems.

\subsubsection{Agile Chip Generation}
Prior efforts have considered chip generation from \#pragma annotated C code, mapping execution onto precomputed hardware primitives for both ASIC FPGA and spatial accelerators~\cite{dsagen, overgen}.
Furthermore, recent works have extended generation to neuromorphic computing, targeting TNN models and architecture~\cite{vellaisamy2024tnngen}.
These frameworks limit themselves to specific accelerators or architecture and thus are not extendable to emerging systems.

\subsection{Limitations of This Work}
\name flattens the computational graph into static events and models their performance by capturing sub-event interactions.
Performance models can be either written manually by users for emerging systems (e.g., superconducting, neuromorphic) or generated by profiling tools or compilers for mature systems (e.g., CPUs and GPUs).
Expertise is required to design circuit-level databases for emerging systems, which can be reused by architecture and system innovations.
With mature system stacks, the workflow can be automated for a specific target or stack, something we leave for future works.
\section{Conclusion}
\label{sec:conclusion}


In the golden age of computer architecture, new systems are rapidly expanding the design space, creating a need for fast exploration.
We propose \name to unify representations across system stacks and separate performance models and cost modeling, enabling flexible, tightly coupled cost co-simulation.
Building on these concepts, \thiswork enhances usability through a front-end interface that automatically generates and constrains design spaces.
Moreover, \thiswork provides scope-based metric retrieval and aggregation exposure, improving programmability and explainability for cross-stack cost modeling.
Extensive case studies demonstrate \name's usability in cross-stack system representation while showcasing its applicability to emerging and conventional systems.



\bibliographystyle{IEEEtran}
\bibliography{refs}

\end{document}